\documentclass[orivec]{llncs}
\usepackage{amssymb}
\usepackage{marvosym}
\input{epsf}
\usepackage{amsmath}
\usepackage{color}
\usepackage{graphics}
\usepackage{dsfont}   
\usepackage{stmaryrd} 
\usepackage{multicol}

\newcommand{\var}{{\mathcal{V}}}

\newcommand{\anal}{\textsf{IA}}

\newcommand{\PBES}{\mbox{\sc Pbes}}
\newcommand{\PBESs}{\mbox{{\sc Pbes}s}}

\newcommand{\ABC}{\mbox{\sc Abc}}
\newcommand{\ANNOTATOR}{\mbox{\sc Annotator}}
\newcommand{\VERISOFT}{\mbox{\sc Verisoft}}
\newcommand{\BANDERA}{\mbox{\sc Bandera}}
\newcommand{\JPF}{\mbox{\sc Jpf}}
\newcommand{\SPIN}{\mbox{\sc Spin}}
\newcommand{\FEAVER}{\mbox{\sc FeaVer}}
\newcommand{\SOCKETMC}{\mbox{\sc SocketMC}}

\newcommand{\APIs}{\mbox{{\sc Api}s}}
\newcommand{\BX}{\mbox{\boldmath $X$}}

\newcommand{\CTL}{\mbox{\sc Ctl}}

\newcommand{\PDL}{\mbox{\sc Pdl}}
\newcommand{\ACTL}{\mbox{\sc Actl}}

\newcommand{\CAESAR}{\mbox{\sc C{\ae}sar}}

\newcommand{\OPENCAESAR}{\mbox{\sc Open/C{\ae}sar}}
\newcommand{\EVALUATOR}{\mbox{\sc Evaluator}}

\newcommand{\IA}{\mbox{\sc Influence\_Analysis}}
\newcommand{\CADP}{\mbox{\sc Cadp}}
\newcommand{\CAESARSOLVE}{\mbox{\sc C{\ae}sar\_Solve}}
\newcommand{\BES}{\mbox{\sc Bes}}
\newcommand{\BESs}{\mbox{{\sc Bes}s}}
\newcommand{\LTS}{\mbox{\sc Lts}}
\newcommand{\LTSs}{\mbox{{\sc Lts}s}}
\newcommand{\MES}{\mbox{\sc Mes}}

\newcommand{\MCL}{$\mu$-calculus}
\newcommand{\PMES}{\mbox{\sc Pmes}}

\newcommand{\FALSE}{\mbox{\sf false}}
\newcommand{\TRUE}{\mbox{\sf true}}
\newcommand{\dia}{\left<}
\newcommand{\mond}{\right>}
\newcommand{\brac}{\left[}
\newcommand{\ket}{\right]}
\newcommand{\sem}{[\![}
\newcommand{\antic}{]\!]}
\newcommand{\ARROW}[1]{\stackrel{#1}{\rightarrow}}
\newcommand{\Bool}{{\mathds B}}
\newcommand{\resp}{{\em resp.}}

\newcommand{\If}{\mbox{\bf if}}

\newcommand{\Then}{\mbox{\bf then}}

\newcommand{\Endfor}{\mbox{\bf endfor}}
\newcommand{\Endif}{\mbox{\bf endif}}
\newcommand{\Endwhile}{\mbox{\bf endwhile}}

\newcommand{\While}{\mbox{\bf while}}

\newcommand{\Forall}{\mbox{\bf forall}}

\newcommand{\Do}{\mbox{\bf do}}

\newcommand{\Return}{\mbox{\bf return}}

\begin{document}

\title{\large{Static Analysis using\\Parameterised Boolean Equation Systems}\thanks{This work has been  supported by the  Spanish {\scriptsize MEC}
under grant {\scriptsize TIN2004-7943-C04}.
The second author is also supported by a Lavoisier grant of the French Ministry of Foreign Affairs}}

\author{Mar\'{\i}a del Mar Gallardo \and Christophe Joubert \and Pedro Merino}

\institute{University of M\'alaga\\ Campus de Teatinos s/n,\\ 29071, M\'alaga, Spain\\
{\tt \{gallardo,joubert,pedro\}@lcc.uma.es}}

\date{}

\pagestyle{plain}
\sloppy

\maketitle \thispagestyle{empty}
\begin{abstract}
 The well-known problem of state space explosion in model checking is
 even more critical when applying this technique to programming
 languages, mainly due to the presence of complex data structures.
 One recent and promising approach to deal with this problem is the
 construction of an abstract and correct representation of the global
 program state allowing to match visited states
 during program model exploration. 
 In particular, one powerful method to implement {\em abstract matching}
 is to fill the state vector with a minimal amount of relevant
 variables for each program point. 
 In this paper, we combine the on-the-fly model-checking approach
 (incremental construction of the program state space) and the static
 analysis method called influence analysis (extraction of significant
 variables for each program point) in order to automatically
 construct an abstract matching function. 
 Firstly, we describe the problem as an alternation-free
 value-based $\mu$-calculus formula, whose validity can be checked
 on the program model expressed as a labeled transition system (\LTS). 
 Secondly, we translate the analysis into the local resolution of
 a parameterised boolean equation system (\PBES), whose
 representation enables a more efficient construction of the resulting
 abstract matching function.
 Finally, we show how our proposal may be elegantly integrated 
 into \CADP, a generic framework for both the design and analysis of 
 distributed systems and the development of verification tools. 
\end{abstract}

\section{Introduction}

 One of the most exciting challenges in the model checking community
 is to apply automatic reachability based verification to standard
 programming languages. Actually, there are many ongoing projects
 oriented to adapt the results on formal method research to
 languages like Java (see
 \BANDERA~\cite{Hatcliff-Dwyer-Pasareanu-Robby-03} and
 \JPF~\cite{Brat-Havelund-Park-Visser-00}) or C/C++ (see
 \VERISOFT~\cite{Godefroid-05}, \FEAVER~\cite{Holzmann-Smith-01} or
 \SOCKETMC~\cite{Camara-Gallardo-Merino-06}). As expected, a common
 problem to these approaches is how to deal with the {\em state space
 explosion} problem, resulting from the size of data structures
 employed in real software, which is several orders of magnitude superior to
 the size of models written with formal description techniques.
 
 Abstract interpretation is one well-established solution to automatically 
 construct smaller and sound models, which 
 may be analyzed by model checking tools (see
 \cite{Dams-02,Hatcliff-Dwyer-Pasareanu-Robby-03,Gallardo-Martinez-Merino-Pimentel-04,Ball-Podelski-Rajamani-00}).
 This method, employed in tools like \JPF, \BANDERA\ or
 $\alpha$\SPIN, is partial, because it
 consists in constructing an over-approximation of the program, where
 non-realistic paths are possible.
 Here, we are interested in a more recent approach, which tries to solve 
 the problem using
 precise abstractions. Thanks to a minimal amount of information, such a method 
 explores exactly the paths required for a given property. One technique of particular interest is {\em abstract matching}. 
 It consists in using a function for reducing the state vector by ignoring
 variables, whose values are not
 relevant to check the property. 
 Actually, these variables are
 temporally replaced by their abstractions, allowing to cut the
 exploration paths. 
 Moreover, this approach generates an under-approximation of the whole state 
 space. Thus, it never produces non-realistic paths.
 Holzmann and Joshi were the first in \cite{Holzmann-Joshi-04} to propose the technique, then employed in
 \cite{Pasareanu-Pelanek-Visser-05} and
 \cite{Camara-Gallardo-Merino-06}. One novel contribution in
 \cite{Camara-Gallardo-Merino-06} is the  use of {\em static
 analysis} algorithms to automatically construct  abstraction functions.
 The method makes use of the property
 to be analyzed, and in practice, it is based on computing the influence
 graph for each program variable. 
 
 In this paper, we intend to automatically construct abstract matching
 functions by performing the influence analysis described in
 \cite{Camara-Gallardo-Merino-06} using
 model checking techniques.
 The idea of using model checking to implement static analysis was first 
 expressed by Steffen in~\cite{Steffen-91}, who provided a framework to
 characterize data flow analyses as the verification of
 particular modal formulas. Schmidt then extended Steffen's work in~\cite{Schmidt-98} to relate it with abstract interpretation. More recently,
 the tool j\ABC~\cite{Lamprecht-Margaria-Steffen-06}
 put in practice Steffen's proposals in the context of Java programs.
 Our approach is close to these previous works, but rather focus on one 
 specific analysis: {\em influence analysis}. 
 We show how influence
 analysis can be expressed as an alternation-free modal
 $\mu$-calculus formula with data parameters evaluated on a labeled
 transition system (\LTS) expressing the abstracted program behavior.  
 Another interesting contribution of the paper is the encoding of influence
 analysis in terms of Boolean Equation Systems (\BES). \BESs\ allow a
 natural description of numerous verification problems, such as model 
 checking of
 temporal formulas, bisimulation, partial order reduction, horn
 clause resolution, abstract interpretation and conformance test case
 generation~\cite{Joubert-Mateescu-06}. Moreover, \BESs\ are
 efficiently supported by different resolution algorithms in the
 literature, one implementation being the \CAESARSOLVE\
 library~\cite{Mateescu-06}, which is part of the widespread verification
 toolbox \CADP~\cite{Garavel-Lang-Mateescu-02}. This resolution
 library is used by the model checker \EVALUATOR\
 3.5~\cite{Mateescu-06}, but also by bisimulation and partial-order
 reduction tools. In addition, it has recently been extended with
 distributed algorithms, thus allowing an immediate distribution of
 each tool connected to \CAESARSOLVE~\cite{Joubert-Mateescu-06}.
 Hence, our static analysis proposal can directly benefit from this verification
 platform. Parallelly, the \SOCKETMC\ tool is now being rewritten for
 \OPENCAESAR\ (the new tool being called C2\LTS), thus creating a complete set of
 tools to perform the whole cycle towards verification of software
 with abstract matching functions.

 This paper is organized as follows. Section~\ref{sec:influence}
 summarizes the influence analysis algorithms used to construct
 abstract matching functions. Section~\ref{sec:MCL} translates the
 different algorithms into alternation-free \MCL\ formulas with data
 parameters, and explains the limitations of such an approach.
 Section~\ref{sec:PBES} further transforms the problem into \PBES\
 resolutions. Section~\ref{sec:CADP} shows how to experiment the
 different encodings into the verification toolbox \CADP. Finally,
 Section~\ref{sec:conclusion} gives some concluding remarks and
 directions for future work.

\section{Influence analysis for abstract matching} \label{sec:influence}

 As proposed in \cite{Holzmann-Joshi-04}, an abstract matching
 function {\tt f()} should be invoked when it is necessary to compact
 the state vector. In such cases, the abstraction function computes
 abstract representations of the hidden data and copies the result
 onto the state vector.  In \cite{Holzmann-Joshi-04}, the authors do
 not address any particular method to generate {\tt f()}, however
 they present necessary conditions to define sound abstract functions
 that preserve \CTL\ properties.

 In \cite{Camara-Gallardo-Merino-06} a particular method 
 is proposed to construct {\tt f()} in such a way that the
 function be sound and oriented to the property to be checked. This
 method is based on the identification of  variables that {\em
 influence} the verification result from the current state. In
 particular, the authors of \cite{Camara-Gallardo-Merino-06}
 developed the so-called {\em influence analysis} (\anal) to annotate each
 program point $p$ with the set of {\em significant} variables
 $\anal(p)$ needed to correctly analyze a given property. Data flow analysis \anal\ is a
 variant of the classic live variable analysis ({\textsf LV}) that attaches each program point
 with  the set of
 {\em live} variables at this point. The key difference is that $\anal$
 makes use of the property to be checked to determine the set of {\em
 needed} variables. Informally, a variable is needed (w.r.t. \anal), if its current value may be necessary to
 evaluate the property of interest in the future. Thus, at a given
 point, a {\em live} variable (w.r.t. {\textsf LV}) may not be {\em needed}, if its value
 does not influence the evaluation of the property.
 
 For each program point $p$, $\anal(p)$ is iteratively calculated as the
 fixed point of an operator that informally works as follows. Let $\var$ be the set of program variables.
 $\anal$ starts by attaching to $p$ the set $I(p) \subseteq \var$ of variables, which are initially
 needed at $p$. The definition of $I(p)$ depends on the property to be analysed.
 Now, assume that it is
 known that variable $x \in \var$ is needed at point $p$, then
 variable $y \in \var$ {\em influences}  $x$ at $p$,
 if there exists an execution path in the program from
 $p$ to an assignment $x = exp$, and the current value of $y$
 is used to calculate $exp$.
 The notion of {\em influence} is recursive since
 it may be necessary to check if $y$ influences some variable
 appearing in expression $exp$ in order to decide whether $y$ is
 needed at point $p$. As shown in the following sections, a
 consequence of this recursive behaviour is that we need to use {\em
 parameters} when translating \anal\ into \MCL\ formulas or boolean equation
 systems.
 
 Influence analysis is used in a dual manner by
 hiding (abstracting) the variables, which {\em are not} needed at each
 program point, while the rest of variables remains explicit in the
 state vector. Therefore, the {\em best} \anal\ analysis is the one 
 attaching the smallest set of variables to each point.
 
\begin{figure}[htbp]
      \begin{center}
         \input{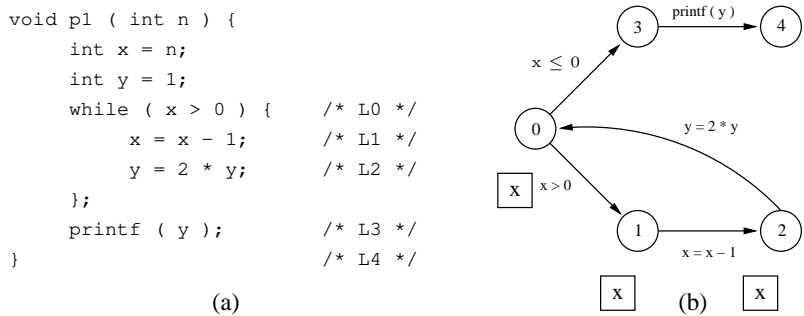}
      \end{center}
   \caption{Example of a C program $p1$ (a) and its control flow
 graph (b)}
   \label{fig:c}
   \end{figure}

 The work in \cite{Camara-Gallardo-Merino-06} describes
 four different influence analyses preserving specific properties.
 The most precise analysis, denoted as  $\anal_1$,
 only preserves information on reachable code. As an example,
 we can consider the C process $p1$, shown in Figure~\ref{fig:c} (a).
 The goal of $\anal_1$ is to determine, in each
 program point (represented as labels $L_0, \cdots, L_4$ in process
 $p1$, and vertices in the corresponding control flow graph
 illustrated in Figure~\ref{fig:c} (b)), which variables will affect
 the program execution flow.

 Figure~\ref{fig:c} (b) shows the intended result of $\anal_1$ for
 $p1$. For this process, the static analysis associates the set
 $\{x\}$ with the labels $L0$, $L1$, and $L2$ (represented in the
 control flow graph as nodes 0, 1 and 2). 
 Hence, if we are interested in knowing whether a
 particular label of process $p1$ is reachable, we only have to
 store variable $x$ at labels $L0$, $L1$, and $L2$. In particular,
 variable $y$ may be completely hidden because its value is not
 relevant for this analysis.
 
 The other variants of $\anal$ extend $\anal_1$ in the following
 way: $\anal_2$ produces bigger sets of variables, but it preserves
 {\em safety properties}. It extends $\anal_1$ considering 
 variables contained in assertions; $\anal_3$ studies the case of
 models with global
 variables; $\anal_4$ is the least precise analysis, but in contrast, it
 preserves {\em liveness properties}. It is based on considering as
 influencing variables all variables appearing in the temporal
 formulas to be verified.
 More details on these influence analyses can be found in
 \cite{Camara-Gallardo-Merino-06}. It is worth noting that they can
 be directly applied to different kinds of modelling and programming
 languages. In particular, in the rest of the paper, we assume
 concurrent systems written in C code.
	     
\section{Mu-calculus model checking for influence analysis}\label{sec:MCL}

This section is devoted to the model-checking of influence analysis over finite \LTSs. We first define the \LTS\ model extracted from the program being statically analysed, next we describe how the influence analysis problem can be translated into the model checking of temporal formulas over the program model, and finally we give the limitations of such an approach.

\subsection{Presentation of the program model}
Influence analysis takes as input a program, or more precisely, a model extracted from it.
In this work, we consider the {\em Labeled Transition System} (\LTS) model, which is suitable for value-passing languages, in particular for concurrent system descriptions.
An \LTS\ is a tuple $\dia S,A,T,s_0 \mond$, where:

\begin{itemize}
\item $S$ is a finite set of states;
\item $A$ is a finite set of actions. An action $a \in A$ is represented as a list $i \vec{w}$, where $i$ identifies the type of actions and $\vec{w}$ is a list of typed values;
\item $T \subseteq S \times A \times S$ is the transition relation. A transition ($s,a,s'$) $\in T$, also noted $s \ARROW{a} s'$, states that the system can move from $s$ to $s'$ by executing action $a$ ($s'$ is an $a$-successor of $s$);
\item $s_0 \in S$ is the initial state. 
\end{itemize}

Furthermore, with respect to the influence analysis problem, we are mainly interested in the set of program variables, that are present in program expressions, such as boolean and assignment expressions.
Thus, we will use only one type of value, for instance the type $Var$ denoting the set of program variables, and we define two types of actions being present in \LTS\ labels:

\begin{itemize}
	\item $BOOL\ \vec{v}$ describes a boolean expression based on the list of variables $\vec{v}$ of type $Var$;
\item $ASSIGN\ v_1.\vec{v}$ describes an assignment expression, where variable $v_1$ of type $Var$ is assigned a value based on variables $\vec{v}$.
\end{itemize}

\example\label{ex:LTS}{
Using the research work of~\cite{Camara-Gallardo-Merino-Sanan-05,Gallardo-Merino-Sanan-06}, which focuses on extracting \LTSs\ out of C programs using well-specified \APIs, we can construct an \LTS\ (see Figure~\ref{fig:LTS}) corresponding to the program presented on Figure~\ref{fig:c}.

   \begin{figure}[htbp]
      \begin{center}
         \input{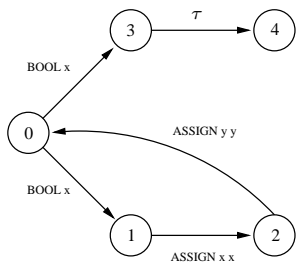}
      \end{center}
   \caption{Example of \LTS\ extended with special actions BOOL and ASSIGN}
   \label{fig:LTS}
   \end{figure}

Its construction results from the control flow analysis of the program together with a labelling of relevant (i.e., $BOOL$ and $ASSIGN$) and invisible (i.e., $\tau$) actions. Moreover, our model splits each action ``$BOOL\ v_1,\ \dots,\ v_j$'' in actions ``$BOOL\ v_i$'' containing only one variable $v_i$, for all $i \in [1,j]$. Similarly, each action ``$ASSIGN\ v_1\ v_2\ \dots\ v_j$'' is split in actions ``$ASSIGN\ v_1\ v_i$'' with two variable parameters only, for all $i \in [ 2, j ]$. 
We can also remark that non-determinism may be introduced artificially (i.e., actions ``$BOOL\ x$'' from state $0$) when creating the \LTS. However, since the unique purpose of such an \LTS\ is to enable influence analysis, all pertinent information for the analysis is kept. Consequently, the static analysis will still have a unique solution.
}

\subsection{Influence analysis using $L_{\mu}^1$ formulas with data parameters}

Modal $\mu$-calculus~\cite{Kozen-83} is an expressive temporal logic based on fixed points, that allows to express a wide range of properties on \LTSs, including those of various other useful logics, such as \PDL~\cite{Fischer-Ladner-79} or \CTL~\cite{Clarke-Emerson-Sistla-86} (as well as its action-based extension \ACTL~\cite{DeNicola-Vaandrager-90}).

The alternation-free fragment of the modal $\mu$-calculus, noted $L_{\mu}^{1}$~\cite{Emerson-Lei-86}, is obtained by forbidding mutual recursive dependencies between minimal and maximal fixed point variables. This logic is of practical usefulness thanks to the existence of linear resolution algorithms in the size of the formula (number of operators) and \LTS\ (number of states and transitions).

In this work, we are interested in the value-based extension of the logic~\cite{Mateescu-98}, which enables the specification of data variables and parameterised fixed point into the temporal formulas. Properties are not restricted to static label description, but they can refer to dynamic values dependent from the system execution. 
Formulas of alternation-free value-based modal $\mu$-calculus are defined by the following grammar (where $X \in \cal{X}$ is a propositional variable, and $\cal{X}$ a set of propositional variables):

\begin{center}
\begin{tabular}{rcl}
$\phi$ & $::=$ & 
$\FALSE
\mid \TRUE
\mid \phi_1 \vee \phi_2
\mid \phi_1 \wedge \phi_2
\mid \dia a \mond \phi
\mid \brac a \ket \phi
\mid X (\vec{e})$ \\
& &
$\mid \mu X (\vec{x}:\vec{t}:=\vec{e}).\phi
\mid \nu X (\vec{x}:\vec{t}:=\vec{e}).\phi$
\end{tabular}
\end{center}

The semantics of a formula $\phi$ over an \LTS\ $M = (S,A,T,s_0)$ denotes the set of states satisfying $\phi$ and it is defined as follows:
boolean operators have their usual definition; possibility operator $\dia a \mond \phi$ (resp. necessity operator $\brac a \ket \phi$) define states from which some (resp. all) transitions labeled by action $a$ lead to states satisfying formula $\phi$; propositional variables $X$ are parameterised by data variables $\vec{e}$; minimal (resp. maximal) fixed point operator $\mu X (\vec{x}:\vec{t}:=\vec{e}).\phi$ (resp. $\nu X (\vec{x}:\vec{t}:=\vec{e}).\phi$) denotes the least (resp. greatest) solution of the fixed point equation $X (\vec{x}:\vec{t}) = \phi$, parameterised by data variables $\vec{x}$ and argument types $\vec{t}$, evaluated with the arguments $\vec{e}$ and interpreted over $2^S$.
On-the-fly model checking determines if the initial state $s_0$ of an \LTS\ satisfies a formula $\phi$ and belongs to the set of states denoted by $\phi$.

Influence analysis is a static program analysis process that it is intended to extract from a specification the set of variables influent on the property evaluation for each program control point.
Although it is a fragment of the data flow analysis problem, which has been shown to be solvable using model checking techniques~\cite{Schmidt-Steffen-98}, namely using the modal \MCL, there doesn't exist to our knowledge a value-based $L_{\mu}^1$ formula encoding the problem of influence analysis.
Our approach is the same in spirit to the one of~\cite{Lamprecht-Margaria-Steffen-06}, where checking a program property corresponds to writing a new formula, evaluating it on the model and extracting from the set of states satisfying the formula, those defining the different program points.
Considering that influence analysis algorithm $\anal_1$ from~\cite{Camara-Gallardo-Merino-06} attaches each program point with the set of variables, whose value is needed to preserve the reachability graph, the resulting value-based $L_{\mu}^1$ formula is:

\begin{center}
\begin{tabular}{rcl}
$\phi_{\anal_1}$ & = $\mu Y (v : Var := x) .$ & 
$(\ \dia BOOL\ v \mond\ \TRUE$ \\ 
& &	      $  \vee\ \dia ASSIGN\ z : Var\ v \mond\ Y(z)$\\
& &	      $  \vee\ \dia \neg (ASSIGN\ v\ z : Var) \mond\ Y(v))$\\
\end{tabular}
\end{center}

Similarly, algorithms $\anal_{2-4}$ can be encoded as a \MCL\ formula.
Since algorithm $\anal_2$ relies on assertions present in the program, it is necessary to extend our \LTS\ with a new type of label:
\begin{itemize}
	\item $ASSERT\ \vec{v}$ describes an assertion composed of variables $\vec{v}$ of type $Var$.
\end{itemize}

$\phi_{\anal_1}$ can naturally be extended by taking into account assertion variables and we obtain the following formula:

\begin{center}
\begin{tabular}{rcl}
$\phi_{\anal_2}$ & = $\mu Y (v : Var := x) .$ & 
$(\ \dia BOOL\ v \mond\ \TRUE$ \\ 
& &	      $  \vee\ \dia ASSERT\ v \mond\ \TRUE$\\
& &	      $  \vee\ \dia ASSIGN\ z : Var\ v \mond\ Y(z)$\\
& &	      $  \vee\ \dia \neg (ASSIGN\ v\ z : Var) \mond\ Y(v))$\\
\end{tabular}
\end{center}

Algorithm $\anal_3$ being an extension of $\anal_1$ and $\anal_2$ considering not only local variables but also global variables, the encoding of the problem as a \MCL\ formula is unchanged and does not need an extra definition.
However, algorithm $\anal_4$ aims at preserving generic temporal properties, and for this purpose, all variables included in such a property have an influence over the program execution.
Since the information contained in temporal properties is external to the program being checked, it will not be accessible in its extracted model, described as \LTS. Hence, checking the influence $\anal_4$ of a variable $x$ at a specific program point is equivalent to, first, test the inclusion of $x$ in the set of variables used in the temporal properties, then, if $x$ is not included, evaluate $\phi_{\anal_4}$ on the \LTS\ as follows:

\begin{center}
\begin{tabular}{rcl}
$\phi_{\anal_4}$ & = $\mu Y (v : Var := x) .$ & 
$(\ \dia BOOL\ v \mond\ \TRUE$ \\ 
& &	      $  \vee\ \dia ASSIGN\ w_i : Var\ v \mond\ \TRUE$\\
& &	      $  \vee\ \dia ASSIGN\ z : Var\ v \mond\ Y(z)$\\
& &	      $  \vee\ \dia \neg (ASSIGN\ v\ z : Var) \mond\ Y(v))$\\
\end{tabular}
\end{center}

The formula $\phi_{\anal_4}$ is an extension of $\phi_{\anal_1}$ with as many modal operations $\dia ASSIGN\ w_i : Var\ v \mond$ as variables $w_i$ present in the external temporal property. Indeed, if a variable $v$ affects the value of $w_i$ in the program, then $v$ is an influent variable itself.

\begin{figure}[htbp]
      \begin{center}
         \input{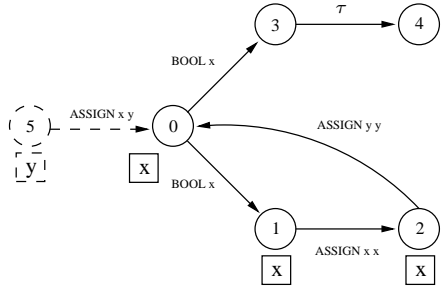}
      \end{center}
   \caption{Example of influence analysis using \MCL\ model checking}
   \label{fig:MCL}
   \end{figure}

\example\label{ex:MCL}{
To illustrate the use of model checking \MCL\ formulas for influence analysis, we can show the result of evaluating $\phi_{\anal_1}$ on the \LTS\ given in Example~\ref{ex:LTS}.
Checking the validity of $\phi_{\anal_1}$ for variable $x$ on state $s_0$ will return $\TRUE$, since there exists boolean expressions (e.g., ``$BOOL\ x$'') involving $x$ reachable from $s_0$. This process can be iterated through all states figuring in the \LTS\ and all variables of the program (i.e., $x$ and $y$), allowing the progressive construction of the list of variables influencing each state (see Figure~\ref{fig:MCL}).
We can remark that only $x$ influences part of the \LTS. Hence, variable $y$ can be totally disregarded without involving any skip of reachable states.
}

\subsection{Limitations of using on-the-fly value-based $L_\mu^1$ model checking}

Instead of iterating through each state, in order to obtain all states satisfying $\phi_{\anal_1}$ for a given variable, it would be more convenient to evaluate only one formula on the whole \LTS, and consequently to extract a subgraph from the original \LTS, containing all states influenced by the specific variable. This could be done by computing $\phi_{\anal_1}$ on the \LTS\ in a backwards manner using a fixed point iteration.
However, this requires the prior computation of the \LTS, and we seek a solution which is suitable for on-the-fly exploration. An adequate \MCL\ formula (for $\anal_1$) would look like the following:
\begin{center}
\begin{tabular}{rcl}
$\phi_{all\anal_1}$ & = $\nu Z .$ & 
$ (\ \phi_{\anal_1}\ \wedge\ \brac\ \TRUE\ \ket\ (\ \neg\ \phi_{\anal_1} \vee\  Z\ )\ )$\\
\end{tabular}
\end{center}
This formula has the same interpretation as $\phi_{\anal_1}$, meaning that its satisfaction on the initial state $s_0$ denotes that the given variable is significant for the initial state. Moreover, the on-the-fly evaluation of $\phi_{all\anal_1}$ on a state satisfying $\phi_{\anal_1}$ requires the recursive evaluation of all its successors that also satisfy $\phi_{\anal_1}$, until all states satisfying $\phi_{\anal_1}$ have been explored. In case of a $\TRUE$ answer, it is then possible to draw a positive diagnostic (example), that only reports the states annotated by $x$ in the Figure~\ref{fig:MCL}. However, this is only true if $x$ never gets assigned a new value. In such a case, this might create {\em holes} in the diagnostic, as can be shown in Figure~\ref{fig:MCL} when adding an artificial new state $s_5$ connected to $s_0$. Evaluating $\phi_{all\anal_1}$ on $s_5$ will return $\FALSE$ for variable $x$, whereas $x$ is influent on states $s_0$, $s_1$ and $s_2$. Standard model checkers are not designed to draw such a diagnostic or a partial one with only states satisfying $\phi_{\anal_1}$. 
Hence, an iteration through all states is necessary to incrementally construct the set of states influenced by a specific variable. 

Working at the level of \MCL\ formulas and standard model checkers, allows to design generic solutions that work not only for influence analysis but, more generally, to many static analyses including data flow analyses~\cite{Lamprecht-Margaria-Steffen-06}. However, using on-the-fly model checking presents limitations such as the reusability of formulas validity for different states given a variable, in order to use previous computations to faster the check of new explored states and variables. In this sense, global model checking would be more appropriate, but is more prone to state space explosion when generating the complete state space and verifying the formula on each of its states. 
Moreover, it would be more convenient to incrementally generate the list of variables that influence each state, in order to define strategies on which variables need to be checked on successor states, thus allowing a gain in the number of computations needed.
To respond to these limitations, a finer-grained encoding of the problem in terms of \PBES\ resolution is preferred and it is described in the following section.

\section{Influence analysis using PBES}
\label{sec:PBES}

This section introduces the Parameterised Boolean Equation System (\PBES) model, and gives a \PBES\ encoding of the influence analysis problem. 

\subsection{Definition of a parameterised boolean equation system}
\label{ssec:def}

A {\em Boolean Equation System} (\BES)~\cite{Andersen-94,Mader-97} is a tuple $B = (x,M_1,\dots,M_n)$, where $x \in \cal{X}$ is a boolean variable, $\cal{X}$ a set of boolean variables, and $M_i$ are equation blocks ($i \in [1,n]$). 
Each block $M_i = \{x_{ij} \stackrel{\sigma_i}{=} op_{ij} \BX_{ij}\}_{j \in [1,m_i]}$ is a set of minimal (resp. maximal) fixed point equations with sign $\sigma_i = \mu$ (resp. $\sigma_i = \nu$). Boolean constants $\FALSE$ and $\TRUE$ abbreviate the empty disjunction $\vee \emptyset$ and the empty conjunction $\wedge \emptyset$ respectively.
A variable $x_{ij}$ depends upon a variable $x_{kl}$ if $x_{kl} \in \BX_{ij}$. A block $M_i$ depends upon a block $M_k$ if some variable of $M_i$ depends upon a variable defined in $M_k$.
A block is {\em closed} if it does not depend upon any other blocks.
A \BES\ is {\em alternation-free} if there are no cyclic dependencies between its blocks. In this case, blocks can be sorted topologically such that a block $M_i$ only depends upon blocks $M_k$ with $k > i$.
The {\em main} variable $x$ must be defined in $M_1$.
In this work, we are interested in the parameterised extension of alternation-free \BES~\cite{Mateescu-98}, called \PBES.
A \PBES\ is a tuple $B$ = ($x\ (\vec{z}:\vec{t}),M_1,\dots,M_n$), where $x \in \cal{X}$ is a boolean variable parameterised by data variables in $\vec{z}$ typed by $\vec{t}$.
Similarly, each block $M_i = \{x_{ij}(\vec{z_{ij}}:\vec{t_{ij}})  \stackrel{\sigma_i}{=} op_{ij} \BX_{ij}\}_{i \in [1,n],\ j \in [1,m_i]}$ is parameterised by data variables in $\vec{z_{ij}}$ typed by $\vec{t_{ij}}$.

The semantics $\sem {\it op} \{ x_1 (\vec{z_{1}}:\vec{t_{1}}), \dots, x_k (\vec{z_{k}}:\vec{t_{k}}) \} \antic \delta$ of a formula ${\it op} \{ x_1 (\vec{z_{1}}:\vec{t_{1}}), \dots, x_k (\vec{z_{k}}:\vec{t_{k}})\}$ w.r.t. $\Bool = \{ \FALSE, \TRUE \}$ and a context $\delta : {\cal X} \rightarrow \Bool$, which must initialize all variables $x_1$, \dots, $x_k$, is the boolean value $\delta (x_1 (\vec{z_{1}}:\vec{t_{1}})) ~ {\it op} ~ \dots ~ {\it op}~ \delta (x_k (\vec{z_{k}}:\vec{t_{k}}))$.
The semantics $\sem M_i \antic \delta$ of a block $M_i$ w.r.t. a context $\delta$ is the $\sigma_i$-fixed point of a vectorial functional ${\Phi_i}_{\delta} : \Bool^{m_i} \rightarrow \Bool^{m_i}$ defined as ${\Phi_i}_{\delta} (b_1, \dots,b_{m_i}) = (\sem {\it op}_{ij} \BX_{ij} \antic (\delta \oslash [ b_1 / x_{i1}, \dots, b_{m_i} / x_{i m_i} ]))_{j \in [ 1, m_i ]}$, where $\delta \oslash [ b_1 /x_{i1}, \dots, b_{m_i} / x_{i m_i} ]$ denotes a context identical to $\delta$ except for variables $x_{i1}, \dots, x_{i m_i}$, which are assigned values $b_1, \dots,b_{m_i}$, respectively.
The semantics of an alternation-free \PBES\ is the value of its main variable $x (\vec{z}:\vec{t})$ given by the solution of $M_1$, i.e., $\delta_1 (x (\vec{z}:\vec{t}))$, where the contexts $\delta_i$ are calculated as follows: $\delta_n = \sem M_n \antic []$ (empty context because $M_n$ is closed), $\delta_i = (\sem M_i \antic \delta_{i+1}) \oslash \delta_{i+1}$ for $i \in [ 1, n-1 ]$ (interpretation of $M_i$ in the context of all blocks $M_k$ with $k > i$).

The {\em local\/} (or {\em on-the-fly\/}) resolution of an alternation-free \PBES\  $B$ = ($x\ (\vec{z}:\vec{t}),M_1,\dots,M_n$) consists in computing the value of $x (\vec{z}:\vec{t})$ by exploring the right-hand sides of the equations in a demand-driven way, without explicitly constructing the blocks.
Several on-the-fly \BES\ resolution algorithms~\cite{Andersen-94,Mader-97} and \PBES\ resolution algorithms~\cite{Mateescu-98,Groote-Willemse-05} are available; here we consider both the approach in~\cite{Mateescu-98}, giving an algorithm to solve alternation-free \PBES, and the approach of~\cite{Andersen-94}, formulating the \BES\ resolution problem in terms of a {\em boolean graph\/} representing the dependencies between boolean variables.

\begin{figure}[htbp]
      \begin{center}
         \input{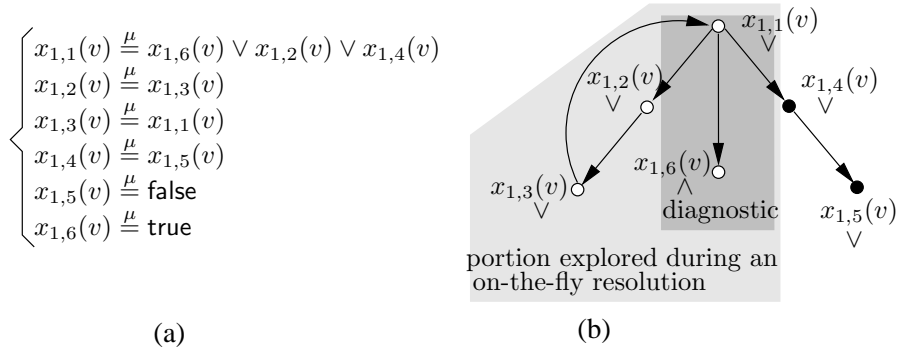}
      \end{center}
   \caption{(a) Example of a parameterised boolean equation system, (b) its boolean graph and the result of an on-the-fly resolution for $x_{1,1}(v)$. Black and white vertices denote \FALSE\ and \TRUE\ variables, respectively.}
   \label{fig:PBES}
   \end{figure}

A boolean graph is a triple $G = (V, E, L)$, where $V = \{x_{ij} (\vec{z_{ij}}:\vec{t_{ij}}) \mid i \in [1,n] \wedge j \in [1,m_i]\}$ is the set of {\em vertices\/} (boolean variables with data parameters), $E: V \rightarrow 2^V,\ E = \{x_{ij} (\vec{z_{ij}}:\vec{t_{ij}}) \rightarrow x_{kl} (\vec{z_{kl}}:\vec{t_{kl}}) \mid x_{kl} \in \BX_{ij}\}$ is the set of {\em edges\/} (dependencies between variables), and $L : V \rightarrow \{\vee, \wedge\},\ L(x_{ij} (\vec{z_{ij}}:\vec{t_{ij}})) = op_{ij}$ is the {\em vertex labeling\/} (disjunctive or conjunctive).
An example of \PBES\ with one block ($i=n=1$) and its associated boolean graph is shown on Figure~\ref{fig:PBES}.

The resolution of variable $x (\vec{z}:\vec{t})$ is performed by a joint forward exploration of the dependencies going out of $x (\vec{z}:\vec{t})$ with a backward propagation of stable variables (whose final value is determined) along dependencies; the resolution terminates either when $x (\vec{z}:\vec{t})$ becomes stable (after propagation of some stable successors) or when the portion of boolean graph reachable from $x (\vec{z}:\vec{t})$ is completely explored. The truth value of $x (\vec{z}:\vec{t})$ can be accompanied by a diagnostic, which provides the minimal amount of information needed for understanding its computed value, as shown in the dark grey area on Figure~\ref{fig:PBES}.

\subsection{Encoding of influence analysis as PBES resolution}
\label{ssec:encoding}

To solve influence analysis using \PBES\ resolution, the first step is to construct an adequate equation system.
Following the approach of~\cite{Mateescu-98}, it is possible to transform the problem of evaluating a value-based alternation-free \MCL\ formula upon an \LTS, into the resolution of a parameterised modal equation system (\PMES) upon the \LTS, by extracting fixed point operators out of the formula.
Starting from $\phi_{\anal_1}$, the resulting \PMES\ contains one block of modal equations and it is given as follows:

\begin{center}
\begin{tabular}{rcl}
$Y (v:Var)$ & $\stackrel{\mu}{=}$ &
$(\ \dia BOOL\ v \mond\ \TRUE$ \\
& &           $  \vee\ \dia ASSIGN\ z : Var\ v \mond\ Y(z)$\\
& &           $  \vee\ \dia \neg (ASSIGN\ v\ z : Var) \mond\ Y(v))$\\
\end{tabular}
\end{center}

Then, to obtain a \PBES\, each modal equation block is converted into a boolean equation block by `projecting' it on each state of the \LTS\ being checked:

\begin{center}
\begin{tabular}{rcl}
	$\{ Y_s (v:Var)$ & $\stackrel{\mu}{=}$ & ${\textstyle \bigvee_{s \ARROW{\it a} s'\ \mid\ a \models BOOL\ v}}\ \TRUE$\\
	& & $\vee {\textstyle \bigvee_{s \ARROW{\it a} s'\ \mid\ a \models ASSIGN\ z\ v}}\ Y_{s'} (z)$\\
	& & $\vee {\textstyle \bigvee_{s \not \ARROW{\it a} s'\ \mid\ a \models ASSIGN\ v\ z}}\ Y_{s'} (v) \}_{s \in S} $\\
\end{tabular}
\end{center}

A boolean variable $Y_s (v)$ is $\TRUE$ iff state $s$ satisfies the propositional variable $Y$ considering variable $v$. Thus, the on-the-fly influence analysis of variable $x$ on the initial state of the \LTS\ amounts to compute the value of variable $Y_{s_0} (x)$. 
The resolution of variable $Y_{s_0} (x)$ on the \LTS\ given in Figure~\ref{fig:LTS} is illustrated on Figure~\ref{fig:PBES}, where variable $x_{1,1}(v)$ corresponds to variable $Y_{s_0} (x)$, and variables $x_{1,j}(v)$ are successors reachable from $Y_{s_0} (x)$, w.r.t. the \PBES\ given above. As shown by the white color, meaning a $\TRUE$ value, of node $x_{1,1}(v)$, variable $x$ is influent on state $s_0$. 
A diagnostic can further be constructed to justify this result by showing a boolean subgraph (in the dark grey area on Figure~\ref{fig:PBES}) containing the variables making $x_{1,1}(v)$ $\TRUE$. For instance, it shows variable $x_{1,2}(v)$, which is a (``$BOOL\ x$'')-successor of $x_{1,1}(v)$, such a transition being the minimal condition for $x$ to be an influence variable.

Generalizing the approach, the influence analysis of all program variables $x$ over all states $s$ contained in the \LTS, can be transformed into an iterative local \PBES\ resolution algorithm.

\begin{figure}
\begin{center}
\begin{tabular}{|p{8cm}|}
\hline
\\
\begin{minipage}{7.5cm}
\begin{tabbing}
AAA\=AA\=AA\=AA\=AA\=\kill
~~1\>\IA\ ($S$,$A$,$T$,$s_0$)\ $\longrightarrow\ S \rightarrow 2^{v(A)}:$\\
~~2\>\>$visited := {s_0};\ explored := \emptyset;$\\
~~3\>\>\While\ $visited \neq \emptyset$ \Do \\
~~4\>\>\>$s := get (visited);\ visited\ := visited \setminus {s};$\\
~~5\>\>\>$explored\ := explored \cup {s};$\\
~~6\>\>\>\Forall\ $v \in var (A)$ \Do \\ 
~~7\>\>\>\>\If\ $solve (Y_s(v))$\ \Then \\
~~8\>\>\>\>\>$d(s)\ := d(s) \cup {v}$\\
~~9\>\>\>\>\Endif \\
~10\>\>\>\Endfor$;$\\
~11\>\>\>\Forall\ $s' \in succ (s) \setminus explored$ \Do \\ 
~12\>\>\>\>$visited\ := visited \cup {s'}$\\
~13\>\>\>\Endfor\\
~14\>\>\Endwhile$;$ \\
~15\>\>\Return\ $d$\\
\end{tabbing}
\end{minipage}
\\
\hline
\end{tabular}
\caption{Influence analysis of \LTS\ using \PBES\ resolution}
\label{algo:IA}
\end{center}
\end{figure}

The function \IA, shown on Figure~\ref{algo:IA}, describes the influence analysis of an \LTS\ $M$ = ($S$,$A$,$T$,$s_0$) using a \PBES\ resolution for each program variable (i.e., $v(A)$) and \LTS\ state. It starts the resolution with initial state $s_0$ (line 2) and iterates through each program variable $v$ (lines 6--10) by constructing and solving the corresponding boolean variable $Y_{s_0}(v)$ (line 7). If the variable $v$ is influent upon the current state, then the set $d(s_0)$ of influence variables for state $s_0$ is increased with variable $v$ (line 8). Next, the process constructs the list of successor states of $s_0$ (lines 11--13), and continues the analysis until all states are explored (line 3).
The result of function \IA\ is the function $d : S \rightarrow 2^{v(A)}$, which returns for each state, the list of variables that are significant.
Such a function $d$ can be further used to automatically construct an abstract matching function stating which variables need to be inserted in the state vector at each program point.
Finally, we can also remark that the algorithm presented on Figure~\ref{algo:IA} can be applied with all influence analysis algorithms ${\anal}_{1-4}$ by using the corresponding \PBES\ encodings when constructing boolean variable $Y_s(v)$ (line 7).

This solution is similar in spirit to the model checking specification in terms of \MCL\ formulas, as it allows to directly provides the desired property as an equation system, whereas it was expressed as a temporal formula in the previous approach.
An important aspect of the method is that influence analysis will require the resolution of only one structure, the parameterised boolean equation system, whereas it needed the resolution of as many \MCL\ formulas as variables being checked, times the number of states in the \LTS. Moreover, the \PBES\ is solved on-the-fly, which means that only the relevant parts of it are computed for each state and each variable. Finally, since a boolean variable $x_{ij}$ defined in $M_i$ may be required several times during the resolution process, it is possible to obtain an efficient overall resolution by using persistent computation results between subsequent resolution calls.

\section{Implementation and experiments}
\label{sec:CADP}

\begin{figure}[htbp]
      \begin{center}
         \input{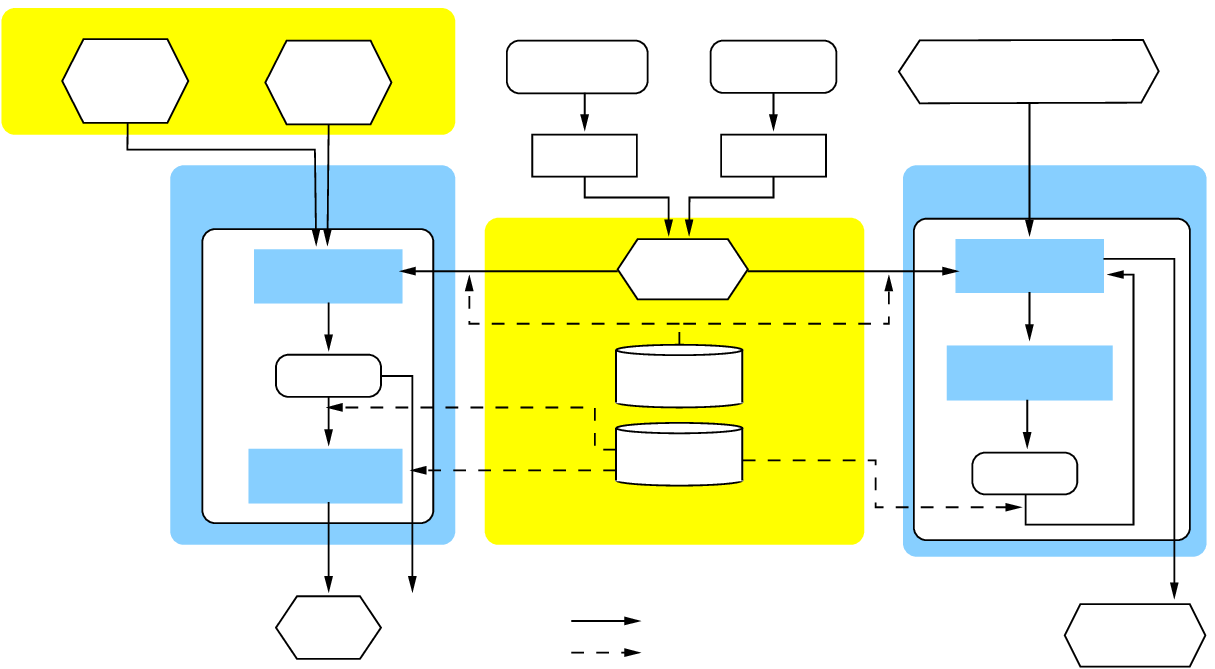}
      \end{center}
   \caption{The on-the-fly tools \EVALUATOR\ and \ANNOTATOR}
   \label{fig:evaluator_annotator}
   \end{figure}

The model checker \EVALUATOR~$3.5$~\cite{Mateescu-06} (see Figure~\ref{fig:evaluator_annotator}) has been developed within \CADP~\cite{Garavel-Lang-Mateescu-02} by using the generic \OPENCAESAR\ environment~\cite{Garavel-98} for on-the-fly exploration of \LTSs.
The static analyser \ANNOTATOR\ on Figure~\ref{fig:evaluator_annotator} is a proposal of tool integrated to \CADP, that applies our \PBES\ approach and follows the same architecture of \EVALUATOR\ 3.5.

\EVALUATOR\ (\resp\ \ANNOTATOR) consists of two parts: a front-end, responsible for encoding the verification of the $L_{\mu}^{1}$ formula (\resp\ the static analysis type) on $\LTS_1$ as a \BES\ (\resp\ \PBES) resolution. \EVALUATOR\ produces also a counterexample by interpreting the diagnostic provided by the \BES\ resolution; and a back-end, responsible of \BES\ (\resp\ \PBES) resolution, playing the role of verification engine.
Both tools are obtained by using, as back-end, algorithms of the \CAESARSOLVE\ library~\cite{Mateescu-06}.
Globally, the approach to on-the-fly model checking (\resp\ static analysis) is both to construct on-the-fly the $\LTS_1$ and corresponding \BES\ (\resp\ \PBES) and to determine the final value of the main variable.

    In the sequel, we present an experimentation with \EVALUATOR\ 3.5 of the influence analysis property $\anal_1$ expressed as a modal equation system (\MES) that is not parameterised, and the structure of \ANNOTATOR\ to achieve the static analysis of an \LTS\ using \PBES\ resolution within \CADP.

\subsection{Experiments with EVALUATOR 3.5}

The current \EVALUATOR\ model checker of \CADP, whose version is 3.5, does not handle data parameters in \MCL\ formulas.
However it is possible to use \EVALUATOR\ 3.5 with the \MCL\ formula $\phi_{\anal_1}$, by transforming it in a parameterless equation system.
This can be done, assuming that the set of program variables $x_i$ is known, by instantiating each call to $Y (x_i)$ into a parameterless propositional variable named $Y_{x_i}$. 
Moreover, to get a more compact representation of the expanded formula, we can use modal equation systems (\MES), which are accepted as input for \EVALUATOR\ 3.5 as {\em .blk} files (option {\em -block}).
Such transformation has already been realized in Section~\ref{ssec:encoding} where the formula $\phi_{\anal_1}$ was expanded into a \PMES.
In order to obtain a resolution complexity linear in the size of the \LTS\ and \PMES, it is necessary to simplify the \PMES, by splitting each right-hand side equation in order to have a single boolean or modal operator~\cite{Mateescu-98}.
Thus, simplifying the \PMES\ $Y$ of Section~\ref{ssec:encoding} leads to the following \PMES :

\begin{center}
\begin{tabular}{rcl}
	$Y_1 (v_1:Var)$ & $\stackrel{\mu}{=}$ & $Y_2 (v_1) \vee Y_3 (v_1)$\\
	$Y_2 (v_2:Var)$ & $\stackrel{\mu}{=}$ & $\dia BOOL\ v_2 \mond\ \TRUE$\\
	$Y_3 (v_3:Var)$ & $\stackrel{\mu}{=}$ & $Y_4 (v_3) \vee Y_5 (v_3)$\\
	$Y_4 (v_4:Var)$ & $\stackrel{\mu}{=}$ & $\dia ASSIGN\ z : Var\ v_4 \mond\ Y(z)$\\
	$Y_5 (v_5:Var)$ & $\stackrel{\mu}{=}$ & $\dia \neg (ASSIGN\ v_5\ z : Var) \mond\ Y(v_5)$\\
\end{tabular}
\end{center}

Next, we transform the simplified \PMES\ in a \MES\ using the parameterless propositional variable $Yj\_{v_i}$.
This \MES\ has a size quadratic w.r.t. the number of influencing variables in the program, but this may be of reasonable size if the number of variables in the program is also not very large.
The {\em .blk} file, for variables $x$ and $y$ in the \LTS\ on Figure~\ref{fig:LTS}, is the following:

\begin{center}
\begin{minipage}{5.25cm}
\fontsize{9}{10}\selectfont {
\texttt{
\begin{tabbing}
A\=AAAAAAAAAAAA\=AAAAAAAA\=AAAA\=AAAA\=AAAA\=AA\=AA\=A\=A\=AAAA\=AAAA\=AAAAA\=\kill
block mu B is \\
\>Y1\_x = Y2\_x or Y3\_x \>\>\>\>\>\>\>\>Y1\_y = Y2\_y or Y3\_y\\
\>Y2\_x = < "BOOL x" > TRUE \>\>\>\>\>\>\>\>Y2\_y = < ``BOOL y'' > TRUE \\
\>Y3\_x = Y4\_x or Y5\_x \>\>\>\>\>\>\>\>Y3\_y = Y4\_y or Y5\_y \\
\>Y4\_x = < "ASSIGN y x" > Y1\_y \>\>\>\>\>\>\>\>Y4\_y = < ``ASSIGN x y'' > Y1\_x \\
\>Y5\_x = < not ("ASSIGN x y") > Y1\_x \>\>\>\>\>\>\>\>Y5\_y = < not (``ASSIGN y x'') > Y1\_y \\
\>\>\>\>\>\>\>\>end block
\end{tabbing}
}}
\end{minipage}
\end{center}

Then, to evaluate the influence of variable $x$ (\resp\ $y$) on the initial state $s_0$, we can use the {\em .blk} clause \verb!eval B:Y1_x! (\resp\ \verb!eval B:Y1_y!), which tells \EVALUATOR\ 3.5 which propositional variable it has to check.
As a consequence, another limit of the method using \EVALUATOR\ 3.5 is that we cannot check the influence property on a state different from the initial state, as \EVALUATOR\ 3.5 will systematically evaluate the \MES\ on the initial state of the considered \LTS.

\subsection{Implementation of an on-the-fly static analyser in CADP}

Instead of using a model checker, we seek a solution that will explicitly manipulate the encoded problem as \PBES, implementing the algorithm given in Figure~\ref{algo:IA}.
This led us to the need of constructing a static analyser in \CADP, based on the \OPENCAESAR\ interface for on-the-fly exploration of \LTS. 

The architecture of such a tool, named \ANNOTATOR, is described on Figure~\ref{fig:evaluator_annotator}. For each visited state in the \LTS, it computes the encoding of the static analysis problem in terms of \PBES\ and solves it upon the state following the algorithm in Figure~\ref{algo:IA}. In the case of influence analysis, the corresponding \PBES, given in Section~\ref{ssec:encoding}, can be projected to the \LTS\ to generate a {\em flat} (i.e., parameterless) \BES, that would be solved by the \CAESARSOLVE\ library. Once the satisfiability of the static property has been computed, the tool can update the definition of a function that returns for each state the result of the analysis (i.e., a set of significant variables in the context of influence analysis). After exploring the entire state space, the annotating function is returned by the tool, and can be further used by other applications, e.g., for abstract matching.

Another important feature of the tool is that both the extracted model (as \LTS) and the \PBES\ can be constructed and explored on-the-fly, thus allowing incremental exploration of only the part of both graphs that is necessary to perform the static analysis.

\section{Conclusion and future work}
\label{sec:conclusion}

Static analysis is a necessary step towards software model checking with abstract matching. Our encodings of the influence analysis problem in terms of alternation-free \MCL\ formulas with data parameters and in terms of \PBES\ resolution enables to automatize the analysis process and to use it in conjunction with on-the-fly verification tools.
To develop robust explicit-state analysis tools, it is necessary to use efficient and generic verification components. Our proposition of on-the-fly static analyser \ANNOTATOR\ goes towards this objective by relying on the generic \OPENCAESAR\ environment~\cite{Garavel-98} for on-the-fly \LTS\ exploration within \CADP~\cite{Garavel-Lang-Mateescu-02} and by using the \BES\ resolution library \CAESARSOLVE~\cite{Mateescu-06}.

We plan to continue our work along several directions. First, we will finish the construction of \ANNOTATOR, as well as the translator C2\LTS\ proposed in~\cite{Gallardo-Merino-Sanan-06} and show the impact of automatic abstract matching on the explored state space size during verification.
Next, we will study the interconnection of both tools integrated into \CADP\ with tools extending \SPIN, such as \SOCKETMC\ and $\alpha$\SPIN~\cite{Gallardo-Martinez-Merino-Pimentel-04}.
Finally, we will seek solutions to other static analysis problems, especially data flow analyses already expressed as \MCL\ formulas in~\cite{Schmidt-98}, by investigating their translation in terms of \PBESs\ resolution.

\paragraph{Acknowledgements.}
We are indebted to Radu Mateescu for its valuable feedback on the possible interaction of our proposal with \CADP\ model checkers.

\end{document}